\documentclass[pra,twocolumn,amsmath,amssymb,aps,nofootinbib,showpacs]{revtex4}

\usepackage{graphicx}				
\usepackage{bm}
\usepackage{subfigure}
\usepackage{comment}
\usepackage{color,linegoal}

\usepackage{soul}
\usepackage[normalem]{ulem}

\newcommand{\epl}{Europhys.\ Lett.\ }

\newcommand{\pr}{Phys.\ Rev.\ }

\newcommand{\jpa}{J. Phys.\ A\ }
\newcommand{\jpb}{J. Phys.\ B\ }
\newcommand{\etal}{{\em et al. }}

\begin{document}

\title{Non-Gaussian pure states and positive Wigner functions}
\author{J. F. Corney and M. K. Olsen}
\date{\today}							

\affiliation{School of Mathematics and Physics, University of Queensland, Brisbane, 
QLD 4072, Australia}

\begin{abstract}

Non-Gaussian correlations in a pure state are inextricably linked with certain non-classical features, such as a non positive-definite Wigner function. In a commonly used simulation technique in ultracold atoms and quantum optics, known as the  truncated Wigner method,  the quantum dynamics is mapped to stochastic trajectories in phase space, governed by a positive approximation to the true Wigner distribution.  The question thus arises: how accurate is this approach in predicting truly non-classical behaviour?
In this article, we benchmark the ability of the truncated Wigner phase-space method to reproduce the non-Gaussian statistics of the single mode anharmonic oscillator.  We find that the this method can reliably predict departures from Gaussian statistics over a wide range of particle numbers, whereas the positive-P representation, which involves no approximations, is limited by rapidly growing statistical uncertainty.  The truncated Wigner function, furthermore, is able to reproduce the non-Gaussian correlations while satisfying the condition for purity.

\end{abstract}


\maketitle

\section{Introduction}
\label{sec:intro}
The Wigner function~\cite{Wigner1}  plays an important role not only in the visualisation and interpretation of quantum states, but also as the basis of calculational techniques.  The Wigner representation maps the quantum-mechanical wavefunction to a phase-space distribution whose marginal distributions give the observed probability distributions for individual variables.  It is tempting to view the Wigner function as a joint probability distribution, but such an interpretation is ruled out by the fact that the Wigner function can take on negative values, in clear distinction to the analogous quantity in a classical system.  Nevertheless, by use of certain approximations, the Wigner function forms the basis of an efficient stochastic approach for quantum dynamics that has been used extensively in quantum optics and ultracold atoms theory. 



There are a number of reasons for investigating how well such phase-space approaches perform in regions where the Wigner function is not positive definite.  First, there is the intrinsic interest of quantum phenomena that have no semiclassical or `hidden variable' explanation.  Second, in a fully quantum treatment, any nonlinear interaction inevitably leads to a non-Gaussian correlations, and thus for a pure state, non-positive definite Wigner functions\cite{Hudson:1974p13087}.  Third, non-Gaussian states and non-Gaussian measurements~\cite{Banaszek} play an important role in quantum information applications such as entanglement distillation~\cite{distill,Heersink:2006p253601} and quantum error correction~\cite{correct}.



In this paper we focus on non-Gaussian pure states  generated by the anharmonic oscillator~\cite{Genoni:2010p13102}, and the positive approximations to their Wigner distributions generated by the truncated Wigner method (TWM)~\cite{Robert,Hardman,Steeletal}.  Since the terms in the Hamiltonian that generate non-Gaussian correlations are the ones that are modified in the resultant Liouville equation by the truncation, there is no {\em a priori} guarantee that any non-Gaussian correlations predicted by the TWM would be accurate.  
%
%
However,  we show that the TWM can indeed reliably predict non-Gaussian behaviour, without introducing any mixedness. 


\section{General overview of phase-space methods}
\label{sec:phasespace}

The idea of mapping quantum states onto equivalent phase-space distributions has been around since the early days of quantum mechanics and the original Wigner 
representation~\cite{Wigner1}, but it took some time for them to be utilised as a basis for simulating quantum dynamics stochastically~\cite{Robert}. To use a phase-space representation for this purpose, we need to be able to sample its dynamics with stochastic trajectories~\cite{CrispinStoch}, which usually requires the distribution to be nonsingular, positive, and governed by a Fokker-Planck equation.  Out of the possible phase-space methods, the two that have emerged as having the greatest practical utility are the (approximate) truncated Wigner and the (exact) positive-P~\cite{P+} methods.  

The truncated Wigner approach has found wide use in studies of both quantum optics  and ultracold atoms.  In quantum optics, it has found to be well suited to treating quantum squeezing and related phenomena~\cite{Granja,JFCfibre}.  However, it fails for example to predict revivals~\cite{Hush2010} that arise from the rephasing of quantum superpositions, and encounters difficulties in calculations of two-time correlations~\cite{arabe}.  For ultracold atoms, the truncated Wigner approach has been used in a wide variety of 
situations~\cite{Steeletal,Sinatra1,Sinatra2,Finn}, and provides the basis for including quantum corrections in the form of vacuum noise to classical field methods~\cite{Ashreview}.  The general rule of thumb is that the average mode occupation should be sufficiently large to avoid ultravoilet divergence issues and the growth of third-order derivative terms in the Liouville equation. Predictions of the atom density (in real or momentum space) have been found to be reliable~\cite{Sarah}, but little attention has been paid to the accuracy of higher-order correlations.

The positive-P representation, by contrast, enables an exact mapping to stochastic equations.  First, since a massive overcompleteness is built into its basis, the positive-P representation allows any quantum state to be mapped to a positive distribution. Second, a wide variety of quantum Hamiltonians lead to Liouville equations for the distribution that contain no derivatives of higher than second order, thus avoiding the need for truncation.  Third, a freedom in the choice of derivatives allows the diffusion-like terms to be cast into a positive-definite form (i.e.\ a Fokker-Planck equation),  which can then be sampled by stochastic trajectories.   However, the resultant noise terms  in the stochastic differential equations are generally multiplicative, which can lead to rapidly growing sampling error. 
 Furthermore, the distribution can develop power-law tails after a certain simulation time that make higher moments undefined, or worse, lead to systematic error if boundary terms no longer vanish~\cite{Gilchrist,Qnoise}. Thus the positive-P method is often limited to short simulation times.

In this paper we begin by presenting a summary of the phase-space descriptions for the anharmonic oscillator dynamics (for both the Wigner and positive-P representations).  In 
section \ref{sec:nonGaussian}, we use cumulants to characterise non-Gaussian behaviour and to benchmark the phase-space methods against exact results.  To check that the truncated Wigner method is not producing a non-Gaussian mixture of Gaussian quantum states, we specifically show in section \ref{sec:puritysection} that non-Gaussian results are obtained without sacrificing the purity of the state \cite{Mandilara:2009p13252}.

\section{Phase-space representations of the anharmonic oscillator}
\label{sec:phasespace}

Besides being about the simplest quantum model that includes nonlinear effects (and hence the possibility of producing non-Gaussian states),  the anharmonic oscillator describes the essential physics of a wide range of coherent nonlinear phenomena.  In  quantum optics, it provides a basic model for the $\chi^{(3)}$ nonlinear response in dielectric materials, such as silica fibre. For degenerate atoms, it models the $s$-wave scattering dominant at  ultracold temperatures.  For both ultracold atoms and pulsed optical experiments, the multimode nature of the quantum fields need to be taken into account for realistic treatments, especially if comparison to experiment is required. Nevertheless, that the single-mode anharmonic oscillator can have exact analytic solutions~\cite{nonKerrCV} gives it an important role in benchmarking methods that can then be generalised to the multimode situation.  This is the end we have in view here, since the phase-space methods we use are easily generalised to, and scale well with, many mode treatments.

The essential idea of the phase-space methods is to map the von Neumann or master  equation for the density operator to an equivalent Liouville equation for a quasiprobability distribution.  If the Liouville equation is in the form of a Fokker-Planck equation, it is simple to sample it with stochastic trajectories in phase space, as long as a positive and nonsingular distribution can be found for the initial quantum state.   Depending on the particular phase-space representation chosen, some approximations or truncations may need to be made to bring the Liouville equation into Fokker-Planck form.

For example, under the anharmonic oscillator Hamiltonian:
\begin{equation}
\hat{H} = \hat{a}^\dagger \hat{a}\hat{a}^\dagger \hat{a}, 
\label{eq:anharmH}
\end{equation}
the Wigner mappings~\cite{Qnoise} lead to a Liouville equation with third-order derivatives:
\begin{eqnarray}
\dot W(\alpha,\alpha^*) &=&  -i \left\{ \frac{\partial}{\partial \alpha} (2|\alpha|^2 -1)\alpha - \frac{\partial}{\partial \alpha} (2|\alpha|^2 -1)\alpha^*  \right. \nonumber\\
&& \left. +   \frac{\partial^3}{{\partial \alpha}^2 \partial \alpha^*} \frac{\alpha}{2}  -  \frac{\partial^3}{\partial \alpha {\partial \alpha^*}^2} \frac{\alpha^*}{2}    \right\}W(\alpha,\alpha^*).
\label{eq:WigL}
\end{eqnarray}
The simplest approach to the third-order terms this equation is to neglect them, to give a Fokker-Planck equation for drift, with no diffusion terms. This approximation is usually justified mathematically by the fact that the third order terms are relatively small for large particle number, and practically because it gives accurate results for many problems.  The corresponding stochastic equation for the amplitude $\alpha$ is then:
\begin{eqnarray}
\frac{d \alpha }{d t} &=& -i\left(2|\alpha|^2-1 \right)\alpha.
\label{eq:alpha}
\end{eqnarray}
Although this equation appears to be deterministic, it retains a stochastic character because the initial conditions  are drawn from appropriate distributions that represent the desired quantum state~\cite{states}.
Note that though the truncation has removed the possibility for an initially positive distribution function to become negative, non-Gaussian behaviour is still possible due to the nonlinear nature of the drift equation.

As an alternative to truncation of the third-order derivatives, one could make use of advanced techniques to map them to stochastic processes~\cite{Wplus,Drummond:2014p13485}.   These third-order processes, however, can have worse stability properties than the positive-P equations~\cite{Wplus}.  A more recent formulation~\cite{Drummond:2014p13485} gives a more compact realisation, but is yet to be fully tested in a Wigner simulation.

For an initial coherent state of amplitude $\alpha_0$, the initial condition is $\alpha(0) =   \alpha_0 + \zeta$,
 where $\zeta$, representing vacuum fluctuations, is a complex-valued white noise term with variance
$\left< \zeta^* \zeta \right>  = \frac{1}{2}$. Stochastic averages in the Wigner representation give the expectation value of symmetrically ordered products of operators.  Therefore powers of the quadrature operators, such as $\langle\hat{X}^2 \rangle $, $\langle\hat{X}^3\rangle$ can be calculated directly. It is also readily seen that the average of $|\alpha|^{2}$ will be equal to $\langle\hat{a}^{\dag}\hat{a}\rangle+1/2$.

For comparison, we will also calculate the correlations using the positive-P representation, which, through the doubling of the effective phase space, allows (i) a positive distribution to be found for any quantum state and (ii) the diffusion term in the Fokker-Planck equation to be made positive definite.  These features, combined with the lack of third- and higher-order terms (for the anharmonic oscillator and similar Hamiltonians)  means that there is an exact mapping to stochastic equations. The one condition for the derivation of the Fokker-Planck equations is that boundary terms must vanish in the partial integration step.  This condition is satisfied at $t=0$ by an appropriate choice of initial distribution, but for some systems may be violated at a later time if power-law tails develop, evidenced by precursor behaviour in the trajectories such as  spiking and large excursions in phase space and the associated rapid increase in sampling error.  For the anharmonic oscillator, it can be shown that such boundary terms will {\em never} appear \cite{Qnoise};  nevertheless, the utility of the method is limited by the rapidly growing sampling error, caused by multiplicative noise and the presence of unstable regions in the enlarged phase space.

The resultant Stratonovich equations for the anharmonic oscillator are:  
\begin{eqnarray}
\frac{d \alpha_1 }{d t} &=& -i(2\alpha_2^*\alpha_1  +  \xi_1(t))\alpha_1\nonumber\\
\frac{d \alpha_2 }{d t} &=& -i(2\alpha_1^*\alpha_2 +  \xi_2(t)) \alpha_2,
\label{eq:Pplus}
\end{eqnarray}
where the multiplicative noises have the correlations:
\begin{equation}
\langle \xi_j(t) \xi_{j'}(t')\rangle = 2i\delta(t-t')\delta_{jj'}.
\label{eq:Pnoise}
\end{equation}
An initial coherent state can be represented by delta function in the positive-P representation, and therefore the initial condition is deterministic:  $\alpha_1(0)  = \alpha_2(0)=   \alpha_0$.  Averages of the positive-P  solutions correspond to normally ordered operator expectation values. For example,  the number of particles $\langle a^\dagger a \rangle$ is given by the average of $\alpha_2^*\alpha_1$.

\section{Non-Gaussian statistics of the anharmonic oscillator }
\label{sec:nonGaussian}

Non-Gaussian states have important roles to play in many quantum information applications~\cite{Genoni:2010p13102}.  For example, they are required for certain distillation schemes~\cite{NGref6}, quantum error correction~\cite{NGref7}, and quantum computation. One method for the production of non-Gaussian statistics is via nonlinear measurements~\cite{NGref8},  but there are limitations in large-scale implementations~\cite{NGref9}.  An alternative approach that avoids such limitations is to make use of relatively simple non-Gaussian sources~\cite{nonKerrCV},  which has the added advantage that standard linear measurements, such as homodyne detection, can be used.  

In order to probe departures from Gaussian behaviour, we use non-zero third and fourth-order cumulants as sufficient conditions for non-Gaussianity:
\begin{eqnarray}
\kappa_3(\theta) &\equiv& \left<\hat{X}^3_\theta\right> - 3\left<\hat{X}_\theta\right>\left<\hat{X}^2_\theta\right> + 2 \left<\hat{X}_\theta\right>^3, \nonumber\\
\kappa_4(\theta) &\equiv& \langle \hat X^4_\theta \rangle + 2\langle \hat X_\theta \rangle^4 - 3\langle \hat X^2_\theta \rangle^2 - 4 \langle \hat X_\theta \rangle \kappa_3(\theta),
\label{eq:skewness}
\end{eqnarray}
where $\hat{X}_\theta = e^{-i\theta}\hat{a} + e^{i \theta}\hat{a}^\dagger$ is the general quadrature variable.  Although these two cumulants do not provide complete information about the non-Gaussianity of the quantum state, which would require the calculation of an infinite number of cumulants, they can be calculated without full knowledge of the density matrix.  The fact that they can be calculated on the basis of just a few moments make them suitable for the stochastic calculation we present here, and possibly also for experimental implementation.

The third-order cumulant is sensitive to the phase of the oscillator, and is zero for symmetric distributions.  Large departures from zero can be obtained for the anharmonic oscillator by using the quadrature in a frame rotating out of phase with the coherent amplitude, i.e.\ $\theta=2N t$, where the initial coherent amplitude is taken to be real.  The fourth-order cumulant is proportional to the to kurtosis, or `peakedness' , of the distribution.  A distribution with positive kurtosis is more peaked (or has fatter tails) than a Gaussian distribution, whereas a negative kurtosis signifies a peak that is broader than a Gaussian. Either of these two cumulants being non-zero is sufficient to demonstrate that the system is non-Gaussian. 

\begin{figure}[htbp] 
  \includegraphics[width=0.9\columnwidth,trim=1cm 6cm 2cm 6cm]
{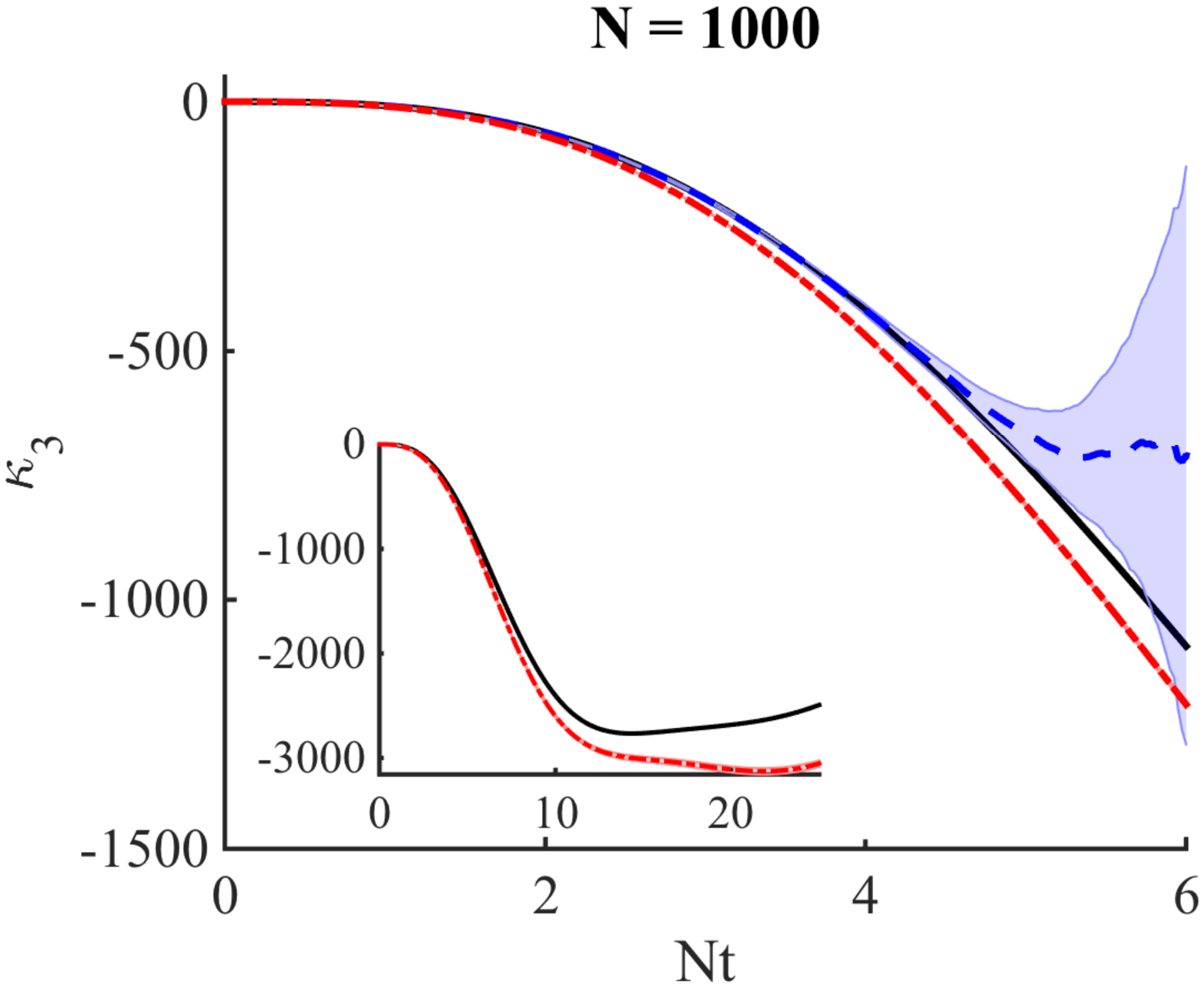} 
   \includegraphics[width=0.9\columnwidth,trim=1cm 6cm 2cm 6cm]
{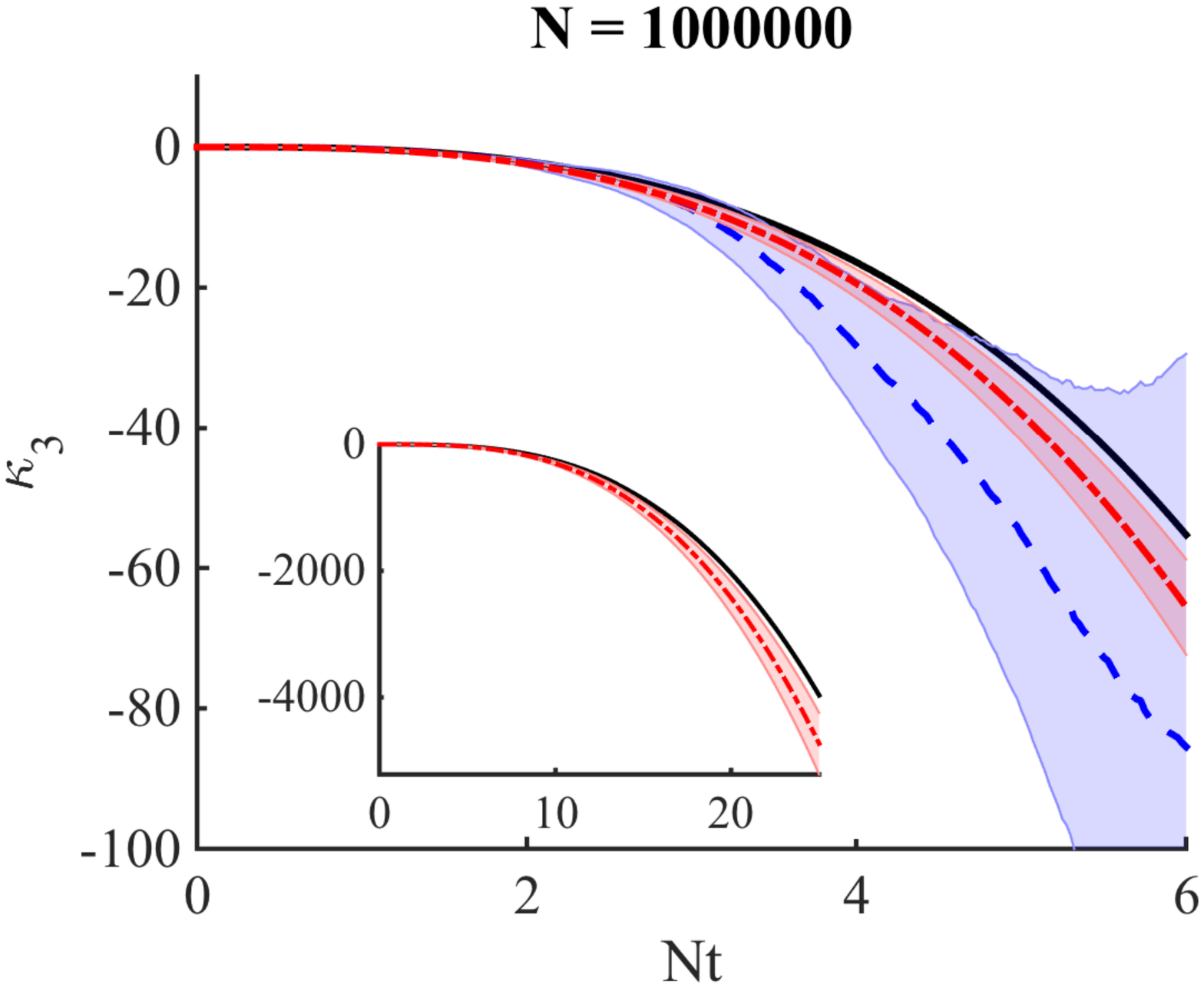} 

  \caption{(Color online) Third-order cumulant of the anharmonic oscillator for $N = 10^3$ (upper plot)   and  $N = 10^6$ particles  (lower plot).  Dot-dashed line (red): truncated Wigner simulations,  dashed line (blue): positive-P simulations, solid line (black): analytic results.   $N_p = 10^8$ stochastic  paths were used for the each simulation, with 
   $\pm \sigma$ estimates of sampling error given by the shading  The insets show the Wigner and analytic results for longer times.}
   \label{fig:cumulants3}
\end{figure}

\begin{figure}[htbp] 
       \includegraphics[width=0.9\columnwidth,trim=1cm 6cm 2cm 6cm]
{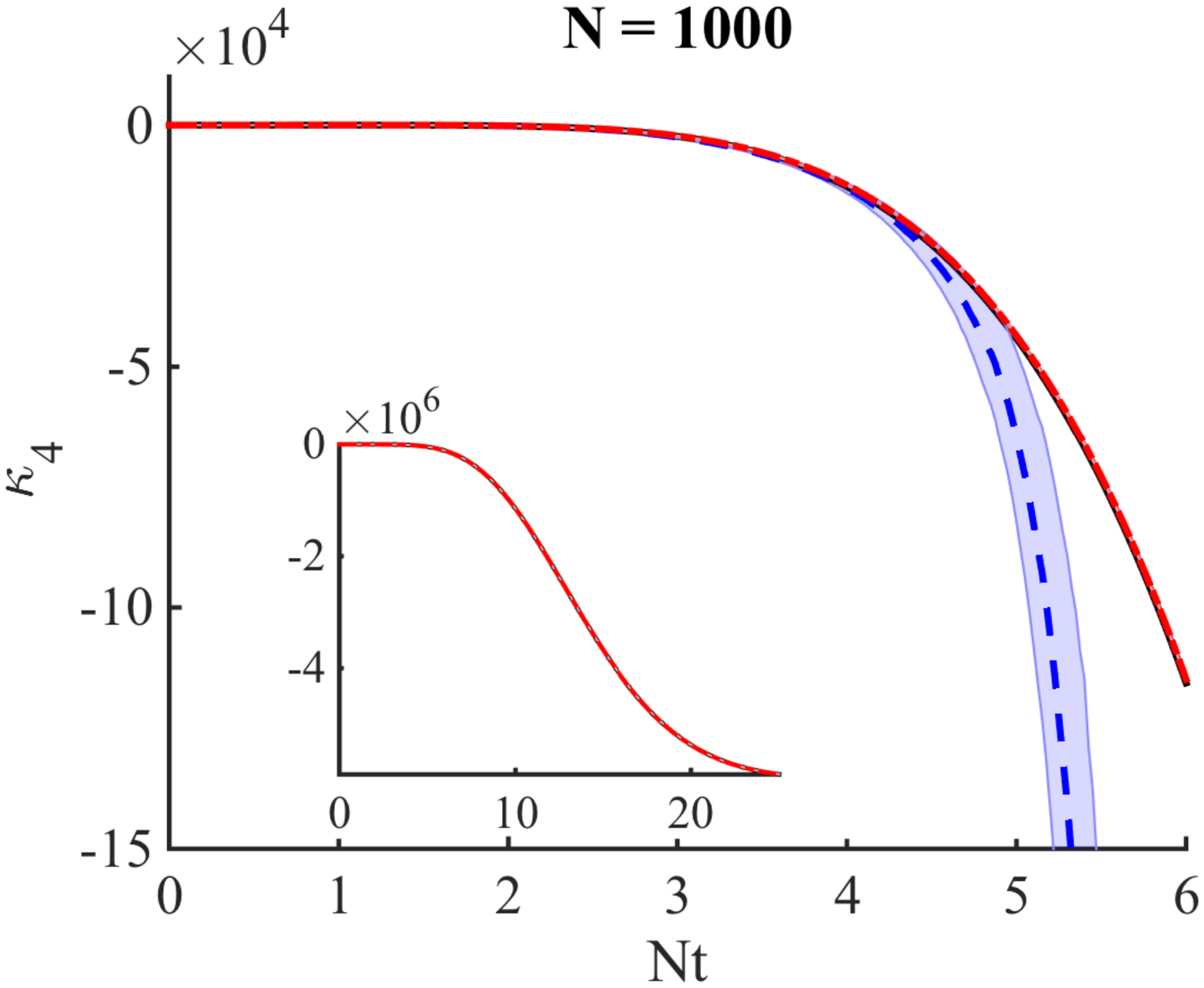} 
    \includegraphics[width=0.9\columnwidth,trim=1cm 6cm 2cm 6cm]
{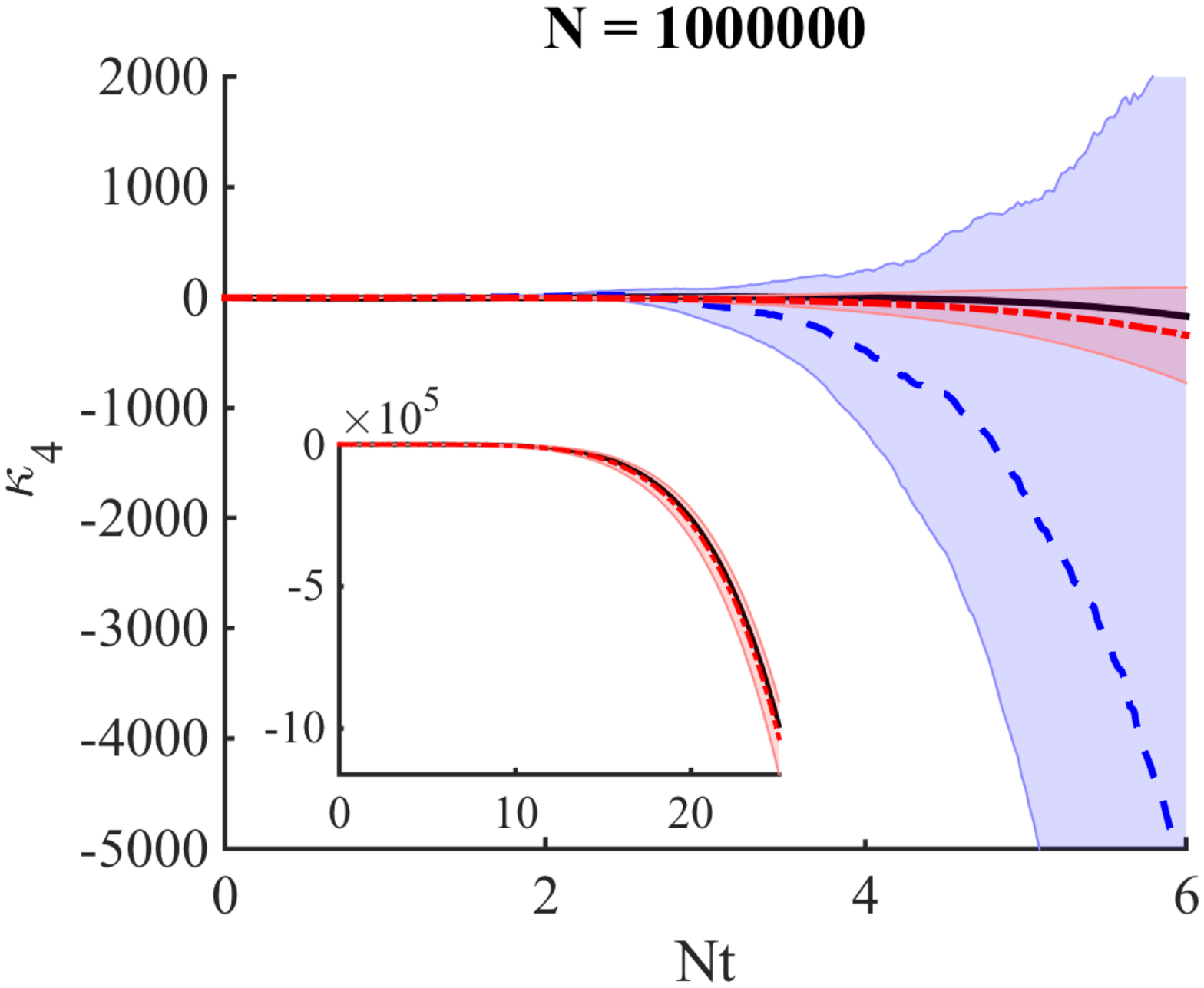} 
   \caption{(Color online) Fourth-order cumulant of the anharmonic oscillator for $N = 10^3$ (upper plot)   and  $N = 10^6$ particles  (lower plot). The symbols and parameters are as in Fig.\ 1.  Note that the positive-P simulation remains within two standard deviations of the analytic curve, even in the upper plot, where the discrepancy temporarily exceeds one standard deviation of the estimated sampling error (between $Nt=5$ and $Nt=6.5$).}
   \label{fig:cumulants4}
\end{figure}

In figures \ref{fig:cumulants3} and \ref{fig:cumulants4} we plot the results of analytic\cite{nonKerrCV} and stochastic calculations of the third and fourth-order cumulants as a function of scaled time $Nt$ for two different particle numbers: $10^3$ (top) and $10^6$ (bottom).  The time-scale shown is sufficient to generate significant squeezing~\cite{Milburn:1986p12570,Tanas:1983p13124}, and the plots in the insets approach the collapse time~\cite{Kirchmair:2013p13126}  $t_c \sim 1/\sqrt{N}$ for the $N = 1000$ case, when the maximum possible non-Gaussian behaviour~\cite{nonKerrCV} is obtained.
We see that the positive-P method (dashed lines) works well for short times, but the rapidly growing sampling error prevents meaningful results past $Nt \approx 5$. 
The stochastic simulations were implemented in XMDS~\cite{Dennis2013201}, with averages being taken over $10^8$ stochastic paths.

For the truncated Wigner method, by contrast, the sampling error is well controlled, even on the longer time-scales shown in the insets ($Nt = 25$).  The agreement with the exact result for the fourth-order cumulant is remarkable and shows that the truncation has had little effect here. The third-order cumulant does reveal some significant systematic error which increases with time, but the discrepancy is of the order of at most 20\% \cite{footnote2}, and the qualitative behaviour is correct. We therefore conclude that the truncated Wigner method can indeed reliably predict departures from Gaussian behaviour, and is thus a useful tool for calculating higher order correlations in nonclassical regimes, at least up to the collapse time. 


\section{Quantumness of the truncated Wigner function}
\label{sec:puritysection}

Since a mixed non-Gaussian state can have a positive Wigner function, it is interesting to consider whether the distribution function corresponding to truncated evolution still obeys the purity condition:
\begin{equation}
\mu \equiv \pi \int W^2(\alpha,\alpha^*) d^2\alpha = 1,
\label{eq:pureza}
\end{equation}
since for a true Wigner function $\mu =  {\rm Tr}[\rho^2]$.

Consider a distribution function that evolves under truncated evolution:
\begin{eqnarray}
\partial_t W({\bm \lambda}) = - \sum_j \partial_{\lambda_j} A_j({\bm \lambda})W({\bm \lambda}),
\end{eqnarray}
where $\bm A$ is the drift vector and $\bm \lambda$ is a vector of the $M$ phase-space variables. Time-evolution of $\mu$ can be shown to be:
\begin{equation}
\dot \mu =- \pi \int \sum_j \left(\frac{\partial A_j({\bm \lambda})}{\partial \lambda_j}\right) W^2({\bm \lambda}) d^M\lambda. 
\end{equation}
In other words, the purity of the initial state is preserved if
\begin{equation}
 \sum_j \left(\frac{\partial A_j({\bm \lambda})}{\partial \lambda_j}\right) = 0,
 \label{eq:purity}
\end{equation}
which is fulfilled for the anharmonic oscillator, for which: 
\begin{equation}
{\bm A} = \left(\begin{array}{c}{-i \alpha^* \alpha^2 }\\{i {\alpha^*}^2\alpha}\end{array}\right).
\end{equation}
More generally, as we show in the appendix, the purity condition for the truncated Wigner is satisfied for any Hamiltonian evolution: $\dot \rho = -i[\hat H,\rho]$.  Thus the non-Gaussian correlations are achieved by a positive approximation to the pure-state Wigner function, rather than through the introduction of mixedness.

\section{Conclusions}

In summary, we have made use of the availability of the analytic solutions of the anharmonic oscillator to benchmark the ability of the truncated Wigner and positive-P techniques to correctly predict departures from Gaussian behaviour, which can be expected to be a challenge for such coherent-state based methods.  We have also explored some of the issues arising from approximating a non-Gaussian Wigner function by a positive distribution.

The results show that the truncated Wigner representation is a suitable tool for the calculation of the third and fourth-order moments, including in regimes where the quantum state is non-Gaussian. Calculating higher-order moments is a challenge for any stochastic method, due to the enhanced sensitivity to statistical fluctuations, and the cumulants in particular, as they involve the near cancellation of large quantities.  Yet, for the truncated Wigner approach, sampling error in these quantities is well-controlled for simulation times long past the time at which the sampling errors in the positive-P become unmanageable.  Furthermore, systematic errors remain small at least up until the collapse time~\cite{footnote1}.


The utility and degree of precision of the truncated Wigner method with this simple example suggests that it will be a suitable method for the calculation of the dynamics of non-Gaussian systems whenever results are required for a time longer than is possible with the positive-P representation. The positive-P, which does not require truncation, will remain the method of choice where sampling error is manageable. The truncated Wigner representation therefore will be an extremely useful tool for quantum information science using complex systems, which tend to have Hilbert spaces too large for other methods to be easily applicable. Given the necessity of non-Gaussian systems for several important quantum information tasks, we feel that many uses of this representation will arise in the future.
 
\section*{Acknowledgments}
This research was supported by the Australian Research Council under the Future Fellowships Program (Grant ID: FT100100515).

\onecolumngrid
\appendix

\section{Phase-space equations of motion}
We here summarise some general results for obtaining phase-space equations of motion, using the standard operator correspondences~\cite{Qnoise}.

\subsection{Positive P representation}

Under the $P$-function mappings, a Hamiltonian $\hat H = \hat{H}(\hat{a}, \hat{a}^\dagger)$ will generate a Liouville equation of the form:
\begin{equation}
\dot P = -i\left\{ H\left(\alpha,\alpha^* - \frac{\partial}{\partial\alpha}\right) - H_r\left(\alpha - \frac{\partial}{\partial{\alpha^*}},\alpha^*\right)  \right\}P,
\end{equation}
where $H_r$ is the reverse-ordered form of $H$.   When the Hamiltonian is given in normally ordered form, the Liouville equation can be written 
\begin{eqnarray}
\dot P(\alpha, \alpha^*)  &=& -i \sum_u \frac{1}{u!}\left\{ \left(-\frac{\partial}{\partial \alpha}\right)^u \left[\frac{\partial^u H(\alpha,\alpha^*)}{{\partial \alpha^*}^u}\right] 
 - \left(-\frac{\partial}{\partial \alpha^*}\right)^u \left[\frac{\partial^u H(\alpha,\alpha^*)}{{\partial \alpha}^u}\right]  \right\}P(\alpha,\alpha^*).\nonumber\\
\end{eqnarray}
If the derivatives higher than second-order vanish, we are able to write down equivalent It\^o stochastic equations:
\begin{eqnarray}
d\alpha &=& -i   \frac{\partial H(\alpha,\alpha^*)}{{\partial \alpha^*}}dt + \sqrt{-i \frac{\partial^2 H(\alpha,\alpha^*)}{{\partial \alpha^*}^2}}dW_1 \nonumber \\
d\alpha^* &=& i   \frac{\partial H(\alpha,\alpha^*)}{{\partial \alpha}}dt + \sqrt{i \frac{\partial^2 H(\alpha,\alpha^*)}{{\partial \alpha}^2}}dW_2 ,
\end{eqnarray}
where $dW_1$ and $dW_2$ are independent Wiener increments.  Note that in the positive-P representation, $\alpha$ and $\alpha^*$ need to be treated as independent variables.  It is only in the ensemble average that they become complex conjugates.  Note also that the conversion of the stochastic equations to Stratonovich form (as we use in the main body of the paper) leads to an additional linear term in the drift.

\subsection{Truncated Wigner representation}

Hamiltonian evolution in the Wigner representation leads to the Liouville equation:
\begin{eqnarray}
\dot W(\alpha,\alpha^*) &=& -i\left\{ H\left( \alpha + \frac{1}{2}\frac{\partial}{\partial \alpha^*}, \alpha^* - \frac{1}{2}  \frac{\partial}{\partial\alpha}\right)
 - H_r\left(\alpha - \frac{1}{2} \frac{\partial}{\partial{\alpha^*}},\alpha^* + \frac{1}{2} \frac{\partial}{\partial\alpha}  \right) \right\}W(\alpha,\alpha^*). \nonumber\\
\end{eqnarray}
If, as above, $H$ is written in normally ordered form, then the Liouville equation can be written in the form
\begin{eqnarray}
\dot W(\alpha,\alpha)&=& -i \sum_{v} \sum_{u} \left( \frac{1}{2} \frac{\partial}{\partial\alpha^*}\right)^{v} \left(\frac{1}{2} \frac{\partial}{\partial\alpha}\right)^{u}  \frac{1}{v!u!} \left\{   (-1)^{u} - (-1)^v\right\} 
 \sum_{w} (-\frac{1}{2} )^{w}  \frac{1}{w!} \left[ \frac{\partial^{2w+u+v}H(\alpha,\alpha^*)}{{\partial \alpha^*}^{u + w }\partial \alpha^{v+w}}\right]W(\alpha,\alpha),
 \end{eqnarray}
from which we can see that any even-order derivative will vanish.  Any Hamiltonian capable of generating non-Gaussian correlations will be higher than quadratic order in the annihilation and creation operators, and hence will lead to third (and possibly higher) order derivatives in the Liouville equation, which will need to be neglected in order to bring it to Fokker-Planck form.  Once this truncation is done, we can write down an equivalent set of drift equations for the phase-space variables:
\begin{eqnarray}
\dot \alpha &=& - i  \sum_{w} (-\frac{1}{2})^{w}\frac{1}{w!} \frac{\partial^{2w+1}H(\alpha,\alpha^*)}{{\partial \alpha^*}^{1 + w }\partial \alpha^{w}} \nonumber \\
\dot \alpha^* &=&  i  \sum_{w} (-\frac{1}{2})^{w}\frac{1}{w!} \frac{\partial^{2w+1}H(\alpha,\alpha^*)}{{\partial \alpha^*}^{ w }\partial \alpha^{1+w}}.
\end{eqnarray}
It is straightforward to see that these drift equations satisfy the purity condition Eq.\ (\ref{eq:purity}).


\begin{thebibliography}{99}
\bibitem{footnote2}  The estimates of sampling error used here may not be totally reliable, particularly for longer times, because they assume (a) Gaussian statistics and (b) that the different moments that go into the cumulant are independent.
\end{thebibliography}
\end{document}